\documentclass{ieeeaccess}
\usepackage{cite}
\usepackage{amsmath,amssymb,amsfonts}
\usepackage{algorithmic}
\usepackage{graphicx}
\usepackage{textcomp}
\def\BibTeX{{\rm B\kern-.05em{\sc i\kern-.025em b}\kern-.08em
    T\kern-.1667em\lower.7ex\hbox{E}\kern-.125emX}}
\usepackage{nicefrac}
\usepackage{siunitx}
\usepackage{array,framed}
\usepackage{booktabs}  
\usepackage{graphics}
\usepackage{xparse} 
\usepackage{xspace}
\usepackage{multirow}
\usepackage{csvsimple}
\usepackage{balance}
\usepackage{listings} 
\usepackage{caption}
\usepackage{float}
\usepackage{braket}

\usepackage{url}
\usepackage[table,xcdraw]{xcolor}

\definecolor{codegreen}{rgb}{0,0.6,0}
\definecolor{codegray}{rgb}{0.5,0.5,0.5}
\definecolor{codepurple}{rgb}{0.58,0,0.82}
\definecolor{backcolour}{rgb}{0.95,0.95,0.92}
\lstdefinestyle{mystyle}{
  backgroundcolor=\color{backcolour}, commentstyle=\color{codegreen},
  keywordstyle=\color{magenta},
  numberstyle=\tiny\color{codegray},
  stringstyle=\color{codepurple},
  basicstyle=\ttfamily\footnotesize,
  breakatwhitespace=false,         
  breaklines=true,                 
  captionpos=b,                    
  keepspaces=true,                 
  numbers=left,                    
  numbersep=5pt,                  
  showspaces=false,                
  showstringspaces=false,
  showtabs=false,                  
  tabsize=2
}

\lstdefinestyle{PythonOutput}{
  basicstyle=\ttfamily\small\color{codegray},
  numbers=none,
}

\begin{document}
\doi{10.1109/TQE.2023.DOI}

\title{Testing and Debugging Quantum Circuits}

\author{
\IEEEauthorblockN{Sara Ayman Metwalli\IEEEauthorrefmark{1} \IEEEauthorrefmark{3}
and Rodney Van Meter\IEEEauthorrefmark{1}\IEEEauthorrefmark{2}\IEEEauthorrefmark{3}}\\ 
\IEEEauthorblockA{\IEEEauthorrefmark{1}\textit{Graduate School of Media and Governance, Keio University Shonan Fujisawa Campus, Fujisawa, Japan}}
\IEEEauthorblockA{\IEEEauthorrefmark{2}\textit{Faculty of Environment and Information Studies, Keio University Shonan Fujisawa Campus, Fujisawa, Japan}}
\IEEEauthorblockA{\IEEEauthorrefmark{3}\textit{Quantum Computing Center, Keio University, Fujisawa, Japan \\ \{sara, rdv\}@sfc.wide.ad.jp}}
}
%

\corresp{Corresponding author: Sara Ayman Metwalli (email: sara@sfc.wide.ad.jp).}

\begin{abstract}
This paper introduces a process framework for debugging quantum circuits, focusing on three distinct types of circuit blocks: Amplitude Permutation, Phase Modulation, and Amplitude Redistribution circuit blocks. Our research addresses the critical need for specialized debugging approaches tailored to the unique properties of each circuit type. For Amplitude Permutation Circuits, we propose techniques to correct amplitude permutations mimicking classical operations. In phase modulation circuits, our proposed strategy targets the precise calibration of phase alterations essential for quantum computations. The most complex Amplitude Redistribution Circuits demand advanced methods to adjust probability amplitudes. This research bridges a vital gap in current methodologies and lays the groundwork for future advancements in quantum circuit debugging. Our contributions are twofold: We present a comprehensive unit testing tool (Cirquo) and debugging approaches tailored to the unique demands of quantum computing, and we provide empirical evidence of its effectiveness in optimizing quantum circuit performance. This work is a crucial step toward realizing robust quantum computing systems and their applications in various domains
\footnote{Portions of this work appeared in a QCE22 conference paper~\cite{metwalli2022tool}. This work is supported by the MEXT Quantum Leap Flagship Program Grant Number JPMXS0118067285.}.
\end{abstract}



\maketitle

\section{Introduction}
\label{sec:intro}

During the past several decades, quantum computing has moved from an idea in scientists' minds to the actual systems we have today. Today, quantum computers exist on a small and error-prone level~\cite{preskill2018quantum,ladd10:_quantum_computers} known as NISQ (Noisy Intermediate Scale Quantum computers). Through that journey, researchers in maths, physics, and computer science worked together to develop algorithms that can harness the strength quantum computers can provide in the future~\cite{montanaro2015:qualgo-qi}. Even though quantum technology has advanced rapidly, both the hardware and software aspects of quantum computing are still in their early stages. Hence, scalability is one of the main challenges we need to overcome on both the hardware and the software sides. We need bigger computers, fault-tolerant qubits, and stable software engineering approaches to build real-life, valuable applications on quantum computers~\cite{van-meter19:_quant_tweet_zen}.

In classical software, the development process follows a mature cycle. Two critical stages of the cycle are testing and debugging, and maintaining the application. The application's abstract aspects may be tested with formal specifications, pseudocode, modeling tools, etc. Bugs arise from errors in the specification of a program, in translating the specification into code, or, sometimes, from bugs in the tools themselves. Currently, there are many approaches to testing classical software, both formal and informal~\cite{orso2014software, myers2004art, pan2009toward}. 
Approaches such as unit testing, regression testing, continuous integration, and path coverage testing make building and supporting systems as complex as tens of millions of lines of code, such as the Linux kernel, possible~\cite{love2010linux, bissyande2012diagnosys}.

\begin{figure}[htbp]
\centerline{\includegraphics[scale=0.25]{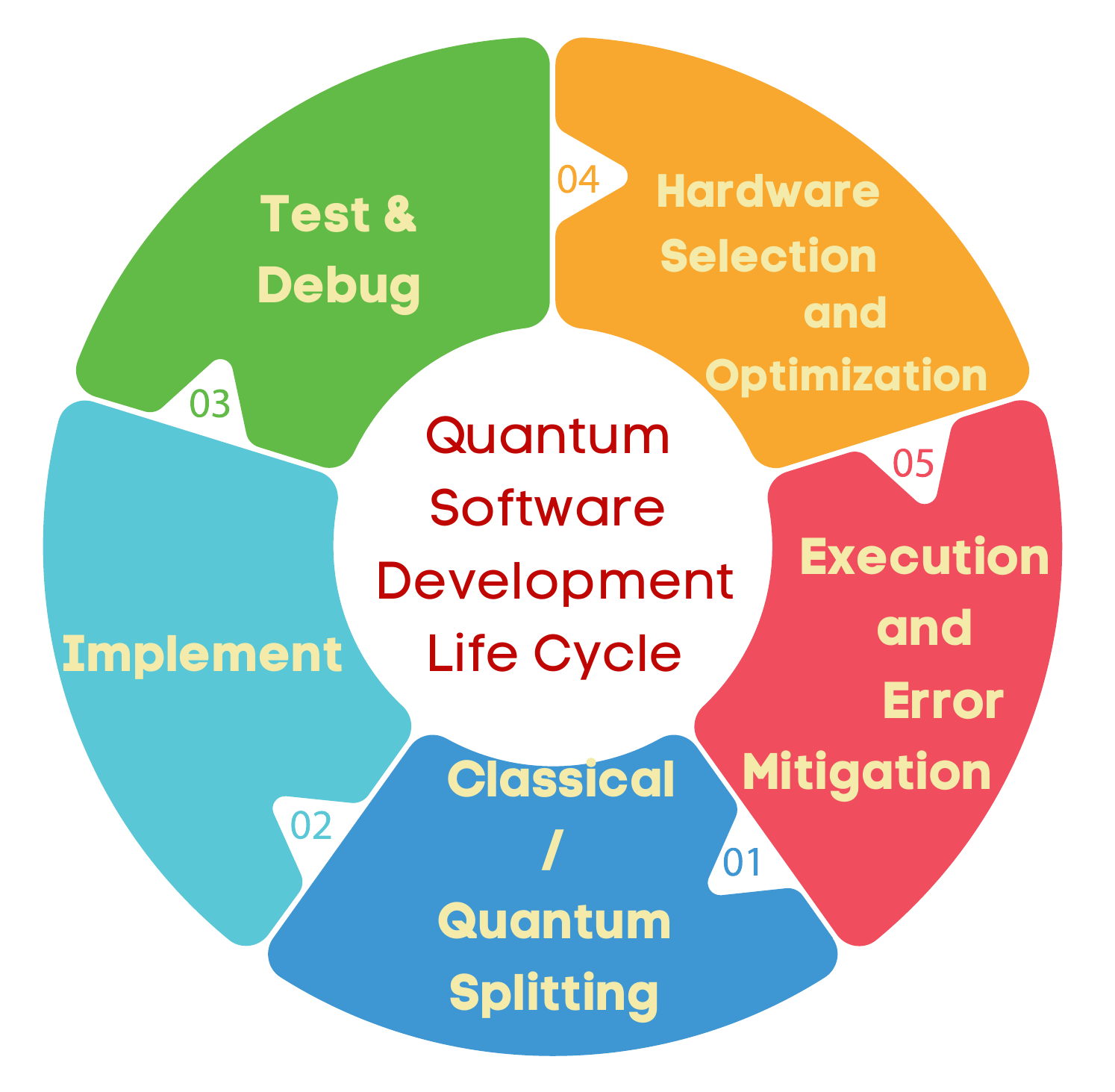}}
\caption{A general life cycle of classical vs. quantum software as described in~\cite{weder2020quantum}.}
\label{cycle}
\end{figure}

Like the classical software development cycle, the quantum software development cycle shown in Figure~\ref{cycle} describes the process of developing software for quantum computers as proposed in~\cite{weder2020quantum}.
Since quantum computers can operate on the \textit{superposition} of values (each with a \textit{complex amplitude}~\cite{sutor19:dancing, preskill:PH-CS219}), the exponential growth in the state space poses a fundamental problem in testing and debugging quantum programs.

When we want to test a quantum circuit, we often have to consider the behavior of all possible inputs {\it as a set}. That exponential growth in the input state space poses fundamental challenges during the testing and debugging process. 
 
The first and perhaps most important challenge is the principle on which quantum algorithms operate. Often, the goal of quantum algorithms is not to find a solution to a problem but rather to build \textit{interference patterns} that amplify the amplitude of correct answers at the expense of the incorrect ones.

Today, developers can use different approaches to transform their ideas and algorithms into quantum programs~\cite{heim2020quantum,gay05:_quant_progr_lang}. If they can develop and implement their algorithms efficiently, leading to small-scale circuits, they can try executing them on actual quantum hardware. Otherwise, they can implement a smaller version of the algorithm and try to extrapolate its behavior for larger instances of the problem. Today's approaches can be categorized into four categories based on the programming model.
\begin{itemize}
    \item \textbf{High-level quantum programming language (QPL)} supporting the developer's quantum intuition such as Silq~\cite{bichsel2020silq}, Q\#~\cite{svore2018qsharp}, and Quipper~\cite{green2013quipper}.  
    \item \textbf{Gate-level programming} In this option, the developer translates their idea into a sequence of gates and then simulates this circuit, visualizes it, or runs it on a hardware device. Developers can use this approach in different ways:
    \begin{itemize}
        \item \textbf{Building the circuit using code} often using a classical-language-supported library or package, such as Qiskit~\cite{aleksandrowicz2019qiskit}, Cirq~\cite{ai2018}, Tket~\cite{sivarajah2020t}, and PyQuil~\cite{pyq} which are Python Packages.
        \item \textbf{Using a drag-and-drop tool} to build the circuit, simulate the results, and view them visually. These tools include QUI~\cite{h2018}, the IBM Circuit Composer~\cite{santos2016ibm}, and Quirk~\cite{gidney}.
        \item \textbf{Using the Quantum Assembly language} or QASM~\cite{cross2017open}.
    \end{itemize}
    \item \textbf{Building the circuit using a low-level approach} such pulses and signals to control the quantum hardware directly. The main example is OpenPulse~\cite{aleksandrowicz2019qiskit}.
\end{itemize}

A summary of some of the most widely used quantum software tools can be seen in Table~\ref{tab:qstools}. All of these tools mainly focus on the current generation of hardware, small programs, and the critical problems of optimization and mapping to specific processors~\cite{chong2017programming, Siraichi:2018:QA:3179541.3168822,nishio20:error-aware, peham2023optimal, wille2023mqt}, as well as on designing and implementing programs for hybrid or adaptive algorithms~\cite{farhi2014quantum, HOGG2000181,mcclean2016theory,peruzzo2014variational, Trugenberger_2002}. 

\begin{table*}[]
\caption{A summary of some widely used current quantum software tools.}
\label{tab:qstools}
\resizebox{\textwidth}{!}{
\begin{tabular}{|
>{\columncolor[HTML]{FFCB2F}}c |c|c|c|c|}
\hline
\cellcolor[HTML]{21A8CF}Name &
  \cellcolor[HTML]{21A8CF}Level &
  \cellcolor[HTML]{21A8CF}Year Released &
  \cellcolor[HTML]{21A8CF}Developer &
  \cellcolor[HTML]{21A8CF}Available For \\ \hline
QuTIP~\cite{johansson2012qutip} &
  Gate-level &
  2012 &
  \begin{tabular}[c]{@{}c@{}}J. R. Johansson, P. D. \\ Nation, and F. Nori\end{tabular} &
  Python Package \\ \hline
Quipper~\cite{green2013quipper} &
  High-level &
  2013 &
  Dalhousie University &
  Quantum PL \\ \hline
Quirk~\cite{gidney} &
  Drag and Drop &
  2016 &
  Craig Gidney &
  \begin{tabular}[c]{@{}c@{}}JavaScript \\ Package\end{tabular} \\ \hline
Qiskit~\cite{aleksandrowicz2019qiskit} &
  Gate-level &
  2017 &
  IBM &
  Python Package \\ \hline
Q\#~\cite{svore2018qsharp} &
  Gate-level &
  2017 &
  Microsoft &
  Quantum PL \\ \hline
QX Simulator~\cite{7927034} &
  QASM-based &
  2017 &
  QuTech &
  QASM \\ \hline
\cellcolor[HTML]{FFCB2F}Ocean SDK~\cite{ocean-dwave-doc} &
  Gate-level &
  2018 &
  D:Wave &
  Python Package \\ \hline
Cirq~\cite{ai2018} &
  Gate-level &
  2018 &
  \begin{tabular}[c]{@{}c@{}}Google\\  (not an official product)\end{tabular} &
  \begin{tabular}[c]{@{}c@{}}Python\\ Package\end{tabular} \\ \hline
Pyquil~\cite{pyq} &
  Gate-level &
  2018 &
  Rigetti &
  Python Package \\ \hline
QUI~\cite{h2018} &
  Drag and Drop &
  2018 &
  \begin{tabular}[c]{@{}c@{}}Hollenberg Group at the \\ University of Melbourne\end{tabular} &
  Quantum PL \\ \hline
PennyLane~\cite{bergholm2018pennylane} &
  Gate-level &
  2018 &
  Xanadu &
  Python Package \\ \hline
Q.js~\cite{quantumjavascriptapp} &
  Drag and Drop &
  2020 &
  Stewart Smith &
  JavaScript Package \\ \hline
Silq~\cite{bichsel2020silq} &
  High-level &
  2020 &
  ETH Zurich &
  Quantum PL \\ \hline
Amazon-Braket~\cite{gonzalez2021cloud} &
  Gate-level &
  2020 &
  Amazon &
  Python Package \\ \hline
TKET~\cite{sivarajah2020t} &
  Gate-level &
  2021 &
  Quantinuum &
  Python Package \\ \hline
Intel SDK~\cite{wu2023comprehensive} &
  Gate-level &
  2023 &
  Intel &
  C++ Library \\ \hline
\end{tabular}%
}
\end{table*}

The responsibility typically falls on developers to thoughtfully plan and manually execute algorithms, as well as to develop tests for checking their outputs using simulators or by reevaluating the underlying mathematics. Yet, this strategy proves impractical when the circuit complexity grows, the current hardware cannot execute them without high error rates, or when classical computers are unable to simulate them effectively.

In this work, we extend upon the tool proposed in~\cite{metwalli2022tool} and offer strategies to debug different quantum circuit types. The ideas of this work are built using Qiskit and Python. Cirquo (the unit-testing package used in this paper) includes a slicer that allows developers to divide their circuits into smaller chunks, categorize, and test them.

The rest of this paper is organized as follows:
\begin{itemize}
    \item Section~\ref{sec:bugs} addresses the possible types of bugs that can occur in quantum programs.
    \item Section~\ref{sec:methodology} introduces the different types of quantum circuits and the characteristics of each of those types.
    \item Section~\ref{sec:test} offers an approach to testing and locating bugs in the different types of circuits.
    \item Section~\ref{sec:dis} wraps up the paper with a discussion of the limitations and challenges of the proposed ideas and approaches and how we can challenge limitations to benefit from these approaches the most.
\end{itemize}

\section{The Sources of Bugs in Quantum Programs}
\label{sec:bugs}

Since understanding the flow of quantum programs and the causes of errors is essential for the ability to debug quantum circuits, researchers focused on categorizing reproducible bugs in quantum programs~\cite{zhao2021bugs4q,campos2021qbugs,luo2022comprehensive}. In these studies, researchers found that bugs in quantum programs can occur for two reasons: the platform used or the developer's implementation. Platform-related errors arise due to a fault in the implementation of the core functionality of the platform, such as deprecation errors, math errors in function implementations, and mistakes in data handling. Paltenghi et al. look in-depth at the bugs introduced by the quantum platform~\cite{pal3527330}. On the other hand, bugs introduced by the developer cover incidents such as the wrong type/ordering of the gates or misuse of the package functions. Both categories will either throw a runtime exception, which is easy to catch and fix, or lead to the wrong answer, making them potentially more challenging to locate and fix.

A quantum bug, at a high level, is an error in creating the correct interference patterns, which can happen due to different reasons depending on the algorithm the programmer is implementing. For example, suppose the programmer is writing an implementation for Grover or Shor's algorithms. In that case, the cause of interference pattern errors can be due to failure in cleaning or detangling the ancilla qubits, a mistake in marking the correct state/s, an improper implementation of the diffusion operator, or, possibly, a combination of all the above. On the other hand, if the programmer implements a quantum chemistry or physics simulation, the errors are "mathematical" due to using the wrong Hamiltonian. Moreover, if the programmer is implementing a variational quantum algorithm (VQA), such as Variational QUantum Eigensolver (VQE) or Quantum Approximate Optimization Algorithm (QAOA), then the source of error could be the wrong ansatz or a mistake in setting the parameters of the classical optimizer. 

Though all these bugs manifest as wrong interference patterns, on a circuit level, they translate to using the wrong gate types/ order, applying the wrong phases, or a combination of both. Section~\ref{sec:test} covers different types of bugs and how Cirquo can assist in locating them.

\section{The Different Types of Quantum Circuits}
\label{sec:methodology}

Classical debugging tools result from decades of research, trials, and developments in software development~\cite{tip1995slicing, kotok1961dec}. The most basic and well-known concept of testing and debugging classical software is program slicing~\cite{xu2005brief}. Program slicing is an approach to dividing a larger program into smaller chunks---slices---to make examining and analyzing it more manageable.

Classically, slices are formed in two ways: manually using breakpoints~\cite{lampson1980processor} or automatic/semiautomatic slicing. In manual slicing, the programmer inserts breakpoints in various locations in the code so that the debugger can divide the code accordingly.

\subsection{Circuit Slicing}
\begin{figure*}[h]
\centerline{\includegraphics[scale=0.3]{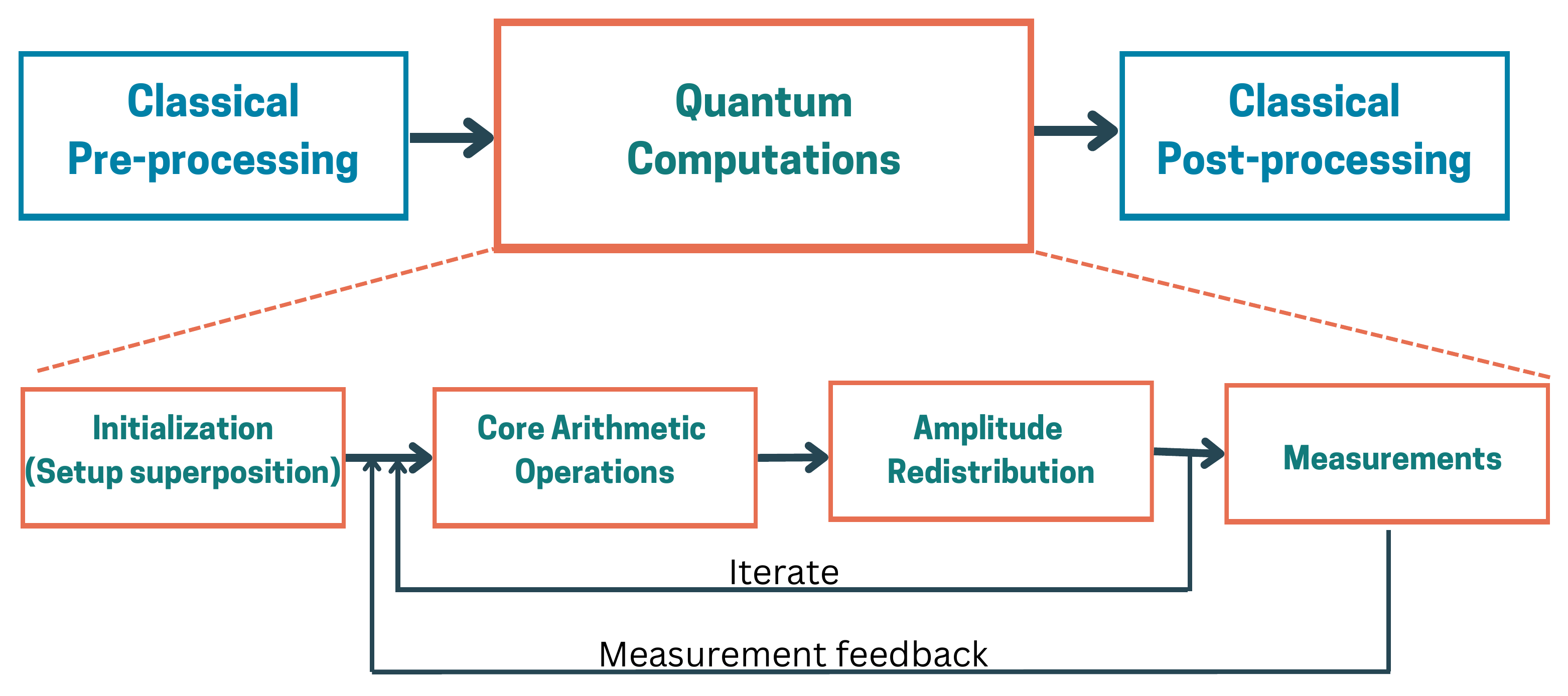}}
\caption{The different steps needed to implement and execute most numerical quantum algorithms assuming mid-circuit measurements.}
\label{algex}
\end{figure*}

Considering the general structure of most quantum algorithms, we notice that they tend to follow a set of steps to solve a problem. Many quantum algorithms start by preparing the working qubits in a specific state or uniform superposition, performing some arithmetic calculations~\cite{takahashi2009quantum}, then redistributing the amplitudes and measuring the results. In the case of some algorithms, we may need to repeat the arithmetic and amplitude distribution steps. Moreover, in most cases
quantum algorithms are accompanied by classical pre-processing or post-processing after the measurement procedure (Figure~\ref{algex}).
For example, Grover's algorithm consists of three main algorithmic steps: preparing the qubits in a uniform superposition, followed by a problem-specific oracle, and then a diffusion operator. In the algorithm, the oracle and diffusion will be repeated multiple times until the amplitude of the answer is predicted to be maximized.
Because of this clear distinction between the algorithmic steps, we can test and examine these steps individually when testing the implementation of an algorithm. Then, put them together to form the target algorithm. 

We can use a similar approach to program slicing classically. We will refer to that as a circuit slicer.
The circuit slicer will divide a large circuit into smaller, executable (or simulatable) subcircuits to demonstrate the functionality of the circuit using both NISQ (Noisy Intermediate-scale Quantum Computers)~\cite{preskill2018quantum} devices and fault-tolerant quantum computers.

In Cirquo's manual circuit slicer, the user inserts breakpoints (in the case of quantum circuits, breakbarriers) in the circuit and then simulates the resultant slices or runs them on an actual device to observe their behavior. 
Considering that quantum circuits are two-dimensional, the slicer can slice the circuits on two axes, the gate axis (vertically) or the qubits/registers axis (horizontally) to remove unused qubits.

\begin{figure*}[htbp]
\centerline{\includegraphics[scale=0.3]{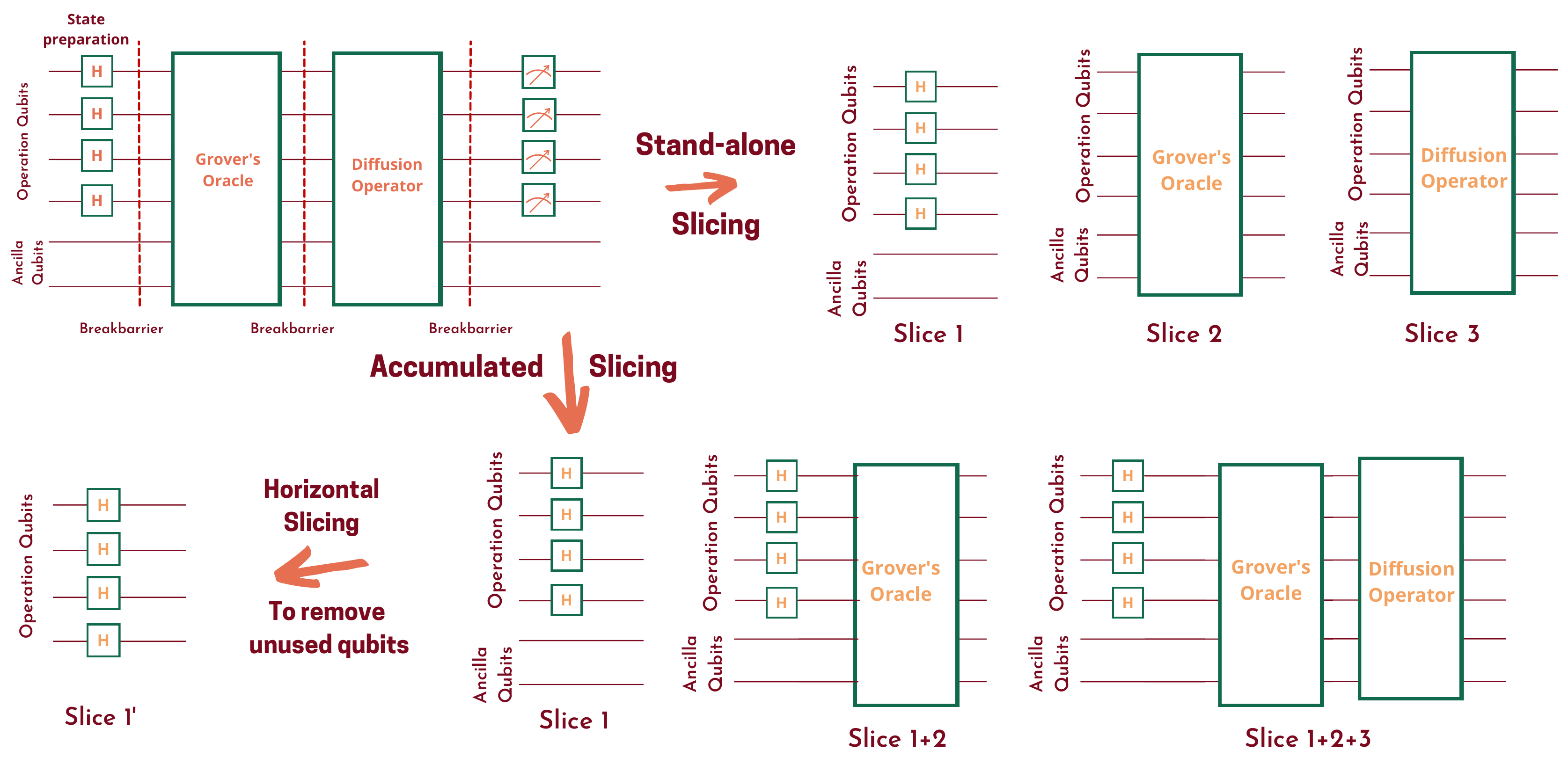}}
\caption{A generic Grover's algorithm circuit is sliced using both stand-alone slicing and accumulated slicing, then one of the slices is horizontally sliced to remove unused qubits.}
\label{slicing}
\end{figure*}

\subsubsection{Vertical Slicing}
To explain the methodology and concept of slicing, let us think of a circuit corresponding to Grover's algorithm~\cite{grover1996fast}. We can use breakbarriers to divide the circuit into slices based on each algorithmic step: the initial state preparation, an oracle, and the diffusion operator. 
To keep things simple, assume that the Grover's algorithm we are slicing consists of one iteration. We will insert two breakbarriers to slice this circuit, one after the state preparation and one after the oracle. This will result in three subcircuits, each performing a specific step in the overall algorithm.

\subsubsection{Horizontal Slicing}
Sometimes, after slicing the circuit vertically, we may end up with a slice that contains some unused qubits. Since our goal of slicing the circuit is to create smaller, simulatable, executable circuits, having unused qubits is redundant. Hence, we can do horizontal slicing to remove these unused qubits from the slice.
The current version of the tool only allows for the automatic slicing of unused qubits. Future expansion will allow users to manually insert horizontal breakbarriers in the case of slices with two independent registers or if the user wants only to include a specific set of qubits.

One main challenge of slicing quantum circuits horizontally is cross-register entanglement before the slice; a way to address that uses the Kronecker product to overcome the challenge of considering the effect of the slice. For example, if we cut only one wire, we get a $4^k$ Kronecker product, where k is the number of qubit wires cut. The math used to develop CutQC~\cite{eddins2022doubling} indicates that the probability of the measurement of an input state $\vert \psi \rangle$ for the unsliced circuit must be equal to the sum of possibilities of the same state for the slices following Equation~\ref{eq:cut}, where $N$ is the number of subcircuits resultant from the slicing.

\begin{equation}
\label{eq:cut}
p(\vert \psi \rangle)=\frac{1}{N} \sum_{i=1}^{4^k} p_{1, i} \otimes ...... \otimes p_{N, i}.
\end{equation}

If the programmer uses a NISQ machine to execute the slices and the original circuit, they need to use an efficient number of shots to achieve good coverage of the possibilities of measuring the different states.

\subsection{Categorizing Circuits}
Considering the construction of quantum algorithms, we can see they consist of three types of building blocks:

\begin{itemize}
\item \textbf{Amplitude-Permutation (AP) Blocks} permute the amplitudes of quantum states. These circuits mimic the operation of reversible classical logic within the quantum realm. Hence, only rearranging the amplitudes associated with the quantum states without redistributing them or altering their phases. An example is a quantum adder or Grover's oracle.
Those blocks are essentially classical reversible logic~\cite{bennett73:reversible,bennett88}.
Mathematically, for set of states $\alpha_{j} \vert j \rangle$, an AP block can be defined as
\begin{equation}
\sum_{j} \alpha_{j} \vert j\rangle \rightarrow \sum_{j} \alpha_{\Pi (j)}\vert j\rangle.
\end{equation}
Where $\Pi(j)$ is a permutation function.
A permutation matrix has exactly one 1 in each row and column. An example of a 2-qubit AP block unitary is
\begin{equation}
A =
\begin{bmatrix}
0 & 1 & 0 & 0\\
1 & 0 & 0 & 0\\
0 & 0 & 1 & 0\\
0 & 0 & 0 & 1
\end{bmatrix}
\end{equation}

\item \textbf{Phase-Modulation (PM) Blocks} Quantum circuits that focus exclusively on altering the phases of quantum states without changing their amplitudes. The primary function of these circuits is to introduce phase shifts based on values of certain qubits.
Mathematically, for set of states $\alpha_{j} \vert j \rangle$, a PM block can be defined as
\begin{equation}
\sum_{j} \alpha_{j}\vert j \rangle \rightarrow \sum_{j} \alpha_{j} e^{i\pi f(j)} \vert j\rangle.
\end{equation}
where $f(j)$ is a function that calculates the phase shift of a state $\theta_{j}$, $f(j)\in \mathcal{R}$.
The unitary of a PM block will be a diagonal matrix $D$ with $D_{jj} = e^{i\pi\theta_{j}}$.

\item \textbf{Amplitude-Redistribution (AR) Blocks} Unlike the Amplitude-Permutation Circuits, these circuits redistribute the amplitudes across various quantum states, thereby harnessing the full potential of quantum superposition and entanglement. An example of an AR block is the Quantum Fourier Transform (QFT).
An AR block contains gates that alter interference patterns and create or destroy superposition. AR blocks can be represented as
\begin{equation}
\sum_{j} \alpha_{j} \vert j\rangle \rightarrow \sum_{j} {\alpha'}_{j}\vert j\rangle.
\end{equation}
Where ${\alpha'}_{j} = \sum_{k} U_{j,k} \alpha_{k}$. Here, $U$ is a unitary matrix applied to the qubits.
\end{itemize}

Often, we can run the AP and PM blocks without the states being in superposition. An example is applying a NOT gate on the least significant qubit (LSQB) in a two-qubit register shown in Figure~\ref{amps}-A (details about reading a Q-sphere are in Appendix~\ref{app2}). This will only permute the amplitude of the states but not affect the superposition or the interference patterns. Similarly, if we apply a T gate on the LSQB of a two-qubit system, it will only affect the phase of the states (Figure~\ref{amps}-B).

\begin{figure}
\centering
\includegraphics[width=\columnwidth]{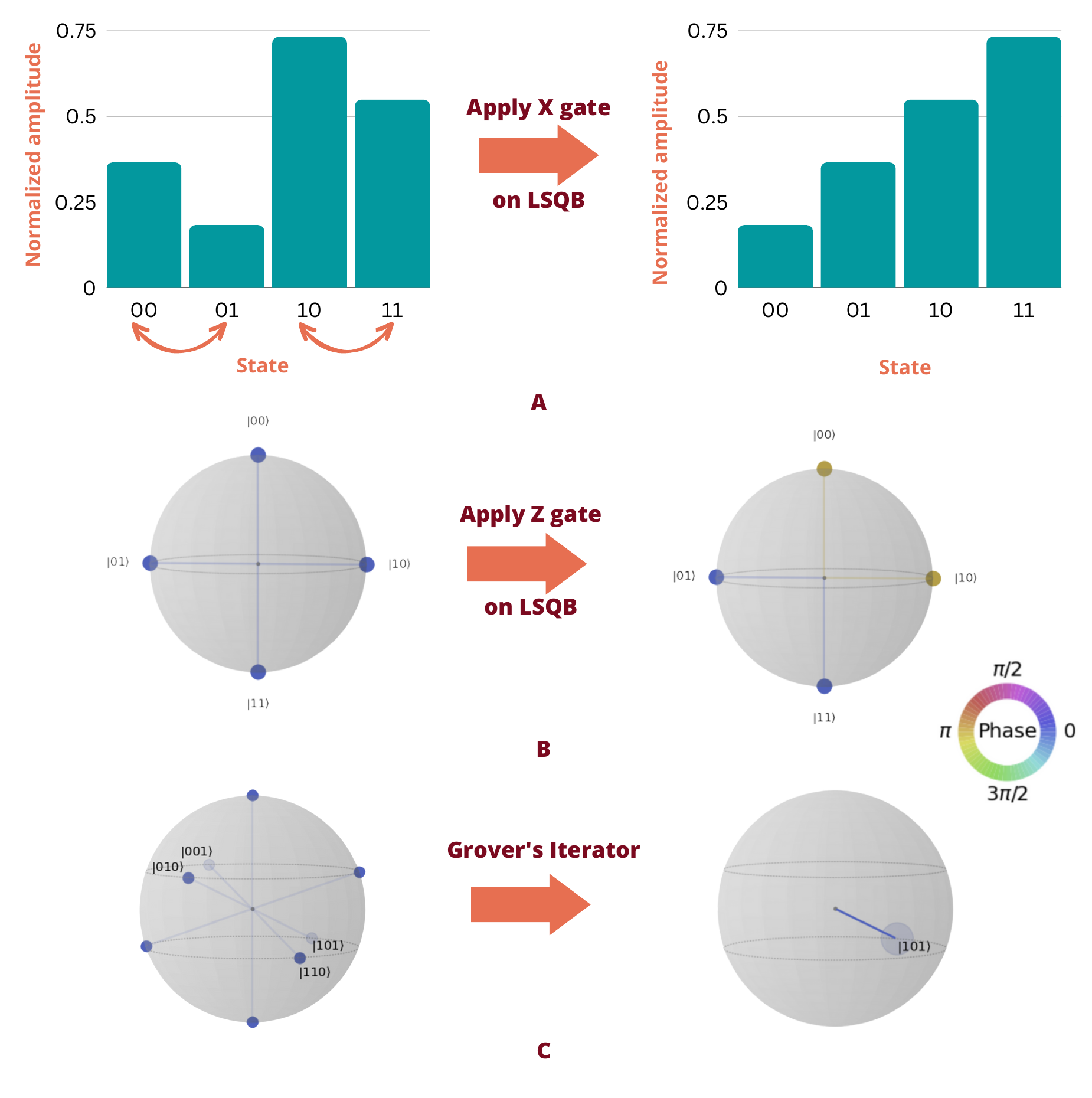}
\caption{Examples of AP, PM, and AR block. A (left) A random amplitude distribution of a 2-qubit state (right) The amplitudes are permuted after applying the NOT gate to the LSQB. B (left) A two-qubit superposition. (right) The phase is modulated after applying a Z gate to the LSQB. C (left) A uniform superposition of 3 qubits (right) The probability amplitude after applying Grover's iterator to mark the correct answer $\vert101\rangle$.}
\label{amps}
\end{figure}

Though Grover's oracle is a PM block, Grover's iterator (the combination of the oracle and the diffusion operator) would be an AR block. Figure~\ref{amps}-C shows the Q-sphere before and after applying Grover's iterator to a 3-qubit circuit.

\section{Testing different circuit types}
\label{sec:test}

To build a framework for testing and debugging quantum programs, first, we must consider the steps of testing a quantum circuit.
Testing a quantum circuit involves several steps, each of which plays a vital role in ensuring the functionality and accuracy of the computations. We can set the needed steps as follows:

\begin{enumerate}
    \item \textbf{Writing Code:} This step involves creating the quantum circuit using the specialized quantum programming languages or quantum packages supported by classical programming languages (Section~\ref{sec:intro}).
    
    \item \textbf{Writing Tests:} Creating test vectors for the circuit and specific parts. This step includes several sub-steps.
    \begin{itemize}
        \item \textit{Slicing the Circuit:} Dividing the circuit into smaller segments for efficient testing.
        \item \textit{Categorizing Slices:} Categorizing the slices based on their functionality within the circuit.
        \item \textit{Adjusting Slice Start/End Points:} Modifying the boundaries of a slice as necessary.
        \item \textit{Developing Test Vectors:} Creating sets of inputs and expected outputs for each slice type.
        \item \textit{Assessing Coverage:} Ensuring that the tests cover all (or most) possible scenarios the slices might encounter.
    \end{itemize}

    \item \textbf{Confidence Interval Selection:}
    Deciding on the statistical confidence level for the test results, which includes:
    \begin{itemize}
        \item \textit{Choosing Optimal Number of Shots:} Determining the number of circuit executions to balance accuracy and computational resource usage is discussed below.
    \end{itemize}
    \item \textbf{Integration:} Resolving issues arise when a slice, which works independently, fails upon integration.
    
    \item \textbf{Running Tests:} Executing the developed test cases against the slices of the quantum circuit and comparing the actual behavior with the expected one. This step can be done on a simulator (if the slice size allows) or an actual device.
    
    \item \textbf{Error Isolation and Additional Testing:} If an error is detected, the problematic slice is isolated and subjected to further focused testing to pinpoint and understand the bug.
\end{enumerate}

These steps provide a systematic approach to this section's testing and debugging process.

As discussed in Section~\ref{sec:methodology}-A.3, we divided quantum circuits into three types with entirely different properties, making the process of testing and debugging them significantly different.
Moreover, testing and debugging the same type will vary depending on its size and the type of instructions it contains. 

AP blocks behave like classical programs; hence, we can use classical approaches when testing them. Therefore, the challenge when testing them would primarily be the difficulty of generating test cases that can provide full coverage. For example, for an adder, we can test a few simple inputs using single amplitudes (no superposition), including overflow cases, and reason by induction for the rest.
Unfortunately, that approach cannot be extended when we deal with AR and PM blocks because they contain quantum properties that are hard to address using traditional testing and debugging techniques. 

Although creating test vectors for AR and PM blocks is not as simple as doing so for the AP blocks, creating test vectors for simple AR and PM blocks should be slightly more straightforward than the complex ones. For example, creating test vectors for a 4-qubit block that contains only 4 Hadamard gates is more straightforward than creating test vectors for a circuit of 20 qubits and 60 different gates.

The size of the block and the types of instructions in it are not the only challenges we face when testing and debugging AR and PM blocks. 

To explain the different strategies for debugging the blocks, we will use the testing functions offered by Cirquo (the complete functionality can be found in Appendix~\ref{app1}).

Before we discuss the different approaches for each type, we need to discuss an approach to testing PM blocks. To do that, we must define a few important terms:
\begin{itemize}
    \item \textbf{$U_{DUT}$}: Device Under Test, which refers to the slice we are testing/ debugging.
    \item \textbf{$U_{TVi}$}: Test Vector $i$, is the circuit corresponding to applying test vector $i$ to $U_{DUT}$. The results of this is $\vert\psi\rangle_{TOi}$, where $\vert\psi\rangle_{TOi} = U_{DUT}U_{TVi}\vert0\rangle$.
    \item \textbf{$U_{EOi}$}: Expected Output $i$, is the expected behavior of $U_{DUT}$ for $TVi$ which is $\vert\psi\rangle_{EOi}$. Because this is specific to a single test vector, it will be substantially simpler than $U_{DUT}$.
\end{itemize}

\begin{figure}
\centering
\includegraphics[width=\columnwidth]{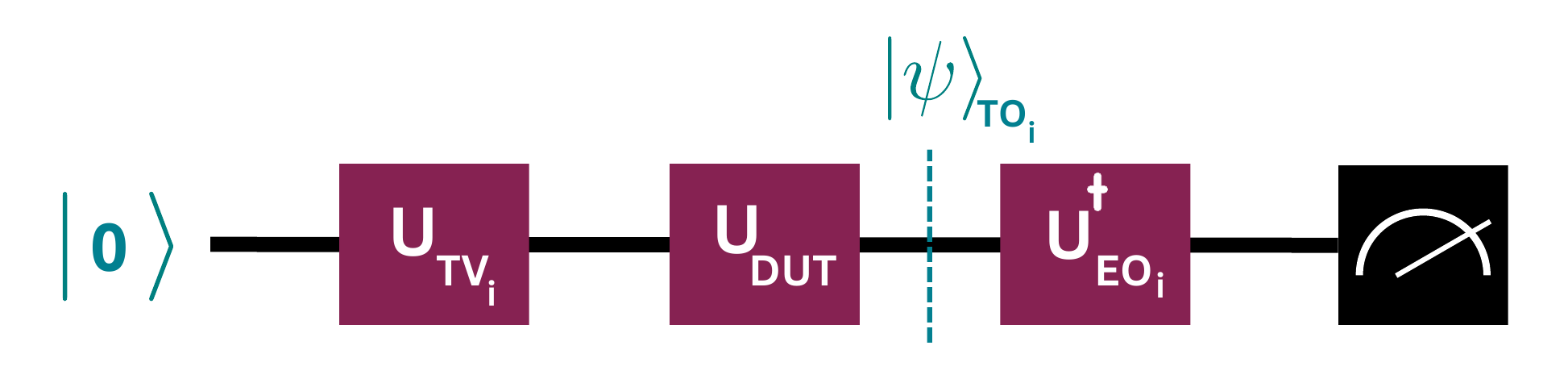}
\caption{One approach to testing quantum circuits, where $U_{DUT}$ is the Device Under Test (full circuit/ slice). $U_{TVi}$ is the circuit corresponding to a test vector $i$. $U_{EOi}$ is the circuit corresponding to the expected behavior for test vector $i$. If $\vert\psi\rangle_{TOi}$ = $\vert\psi\rangle_{EOi}$ the output will be $\vert0\rangle$.}
\label{simtst}
\end{figure}

Figure~\ref{simtst} shows an approach to testing any quantum circuit (regardless of its type). Essentially, what we want to answer is the question: \textit{Does $\vert\psi\rangle_{TOi}$ = $\vert\psi\rangle_{EOi}$}? If the slice we are testing is correct, these two states are equal; if not, we can conclude that something is incorrect in the test circuit and proceed with the debugging process. 
The debugging process will then differ based on the circuit block we target.


\subsection{Testing and debugging Amplitude Permutation (AP) blocks}
\label{subsec:pclass_testing}

As explained in Section~\ref{sec:methodology}, AP blocks are quantum circuits that mimic the behavior of reversible classic logic.

When creating test vectors for an AP block, we follow the same approach for creating test vectors for classical programs. Often, when we create test vectors, the approach follows systematically, covering all possible input combinations and edge cases to verify their functionality thoroughly. 

However, as our target circuits grow, creating test vectors for all possible cases will be increasingly challenging. There, we can focus on some practical cases and edge cases.

\begin{figure}
\centering
\includegraphics[width=\columnwidth]{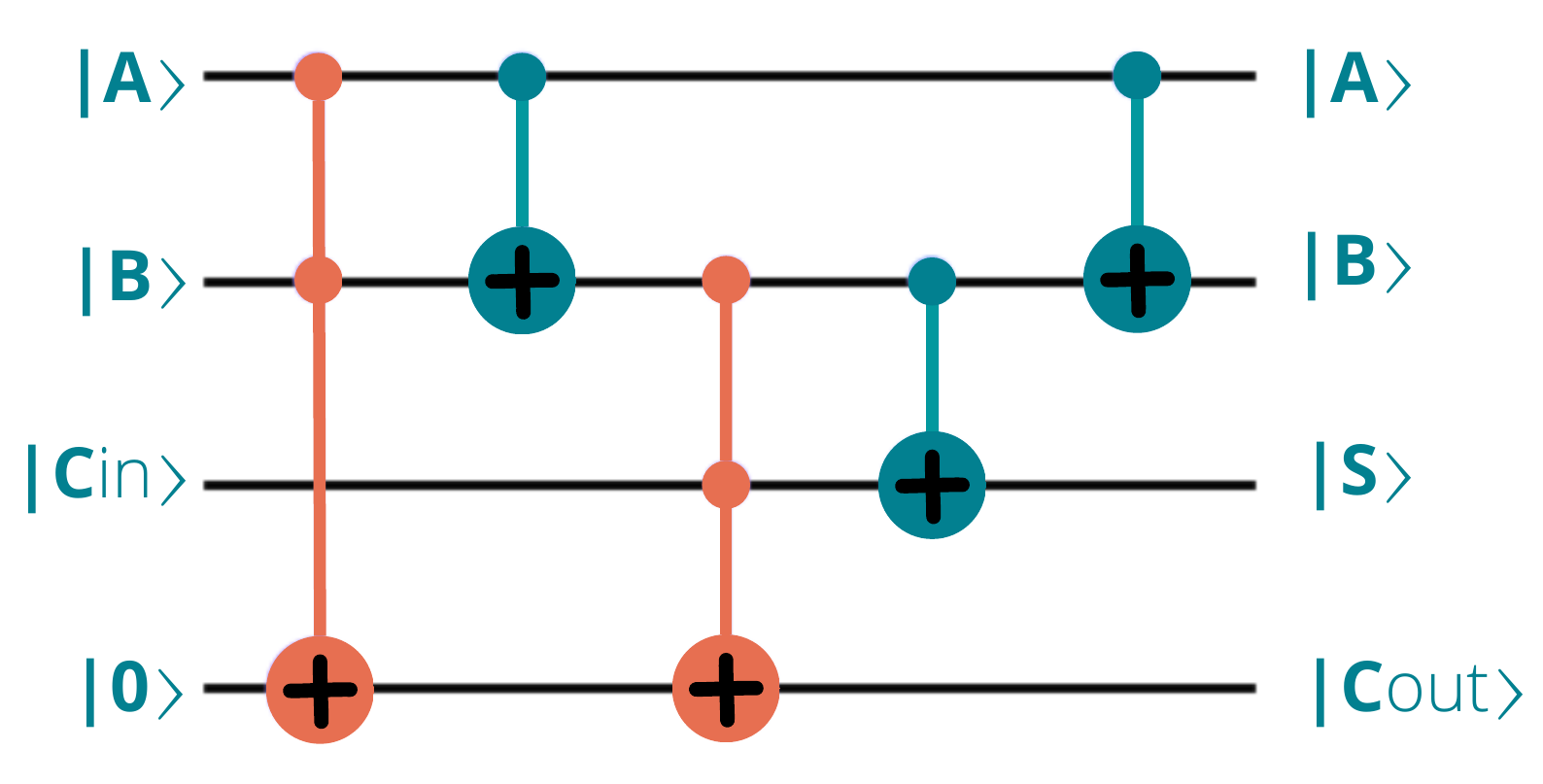}
\caption{The full adder circuit.}
\label{fulladder}
\end{figure}

One example of an AP block is the quantum full adder. The full adder is a 4-qubit system, where the inputs are $\vert A\rangle$, $\vert B\rangle$, $\vert C_{in}\rangle$, and $\vert 0\rangle$. $\vert A\rangle$ and $\vert B\rangle$ are the qubits we wish to add and $\vert C_{in}\rangle$ is the carry-in. The outputs of the full adder are $\vert A\rangle$, $\vert B\rangle$, $\vert S\rangle$, and $\vert C_{out}\rangle$ (Listing~\ref{fulladder_py}, Figure~\ref{fulladder}).

\begin{lstlisting}[language=Python, caption=A simple Python and Qiskit implementation of the full adder., label=fulladder_py]
def Quant_full_adder(qc, in_qbits,zero_qubit):
    qc.ccx(in_qbits[0],in_qbits[1],zero_qubit)
    qc.cx(in_qbits[0],in_qbits[1])
    qc.ccx(in_qbits[1],in_qbits[2],zero_qubit)
    qc.cx(in_qbits[1],in_qbits[2])
    qc.cx(in_qbits[0],in_qbits[1])
    return qc
\end{lstlisting}

The full adder has three inputs $A$, $B$, and $C_{in}$, resulting in a total of $2^3 = 8$ possible input states. Optimally, we can create 8 test vectors for all possible cases. This will ensure every possible state is checked, including edge cases such as all inputs being $0$ or all being $1$ and transitions between these states. We will test our circuit on the five test vectors in Listing~\ref{fulladder_tests}.

\begin{lstlisting}[language=Python, caption=Test vectors for the full adder., label=fulladder_tests]
test_cases = [
    {
        "name": "test 1",
        "input": [1,1,1,0],
        "expected_output": [1,1,1,1]
    },
    {
        "name": "test 2",
        "input": [0,0,0,0],
        "expected_output": [0,0,0,0]
    },
    {
        "name": "test 3",
        "input": [1,0,1,0],
        "expected_output": [1,0,0,1]
    },
    {
        "name": "test 4",
        "input": [0,1,0,0],
        "expected_output": [0,1,1,0]
    },
    {
        "name": "test 5",
        "input": [1,1,0,0],
        "expected_output": [1,1,0,1]
    }
]
\end{lstlisting} 

Running the circuit for these test vectors using Cirquo's \texttt{pClassTester(qc, test\_cases)} results in a PASS for all the tests.

Now, we introduce a simple bug to this program. Assume the programmer inadvertently added an extra Toffoli gate at the end of the circuit \texttt{qc.ccx(in\_qbits[0],in\_qbits[1],zero\_qubit)}.
The first step is running the same tests from before to the program containing the bug. This will lead to two test vectors having a FAIL status, as seen in Listing~\ref{fulladder_res_mis}. 

When we examine the results, we see that the error occurs only when both qubits $\vert A\rangle$ and $\vert B\rangle$ are 1. 

\begin{lstlisting}[language=bash, caption=Failed test results when running the test vectors of the full adder program containing the bug., label=fulladder_res_mis]
Testing test 1:
Result:  FAIL
Input:  [1, 1, 1, 0]
Output:  [1, 1, 1, 0]
Expected Output:  [1, 1, 1, 1]
.
.
Testing test 5:
Result:  FAIL
Input:  [1, 1, 0, 0]
Output:  [1, 1, 0, 0]
Expected Output:  [1, 1, 0, 1]
\end{lstlisting}

In this case, since the error only occurs when the first two qubits are 1, we might guess that they are the control of the gate causing the error. We can locate the multi-qubit gates applied to these two qubits using the \texttt{gateLoc(qc,' cx', qubits=['q[0]','q[1]'])} function, which will lead us to two Toffoli gates, and then we can remove each of them and test the circuit again. Doing so points out that the last Toffoli gate is the source of the error. 

AP blocks are deterministic, so only a single shot is required for each test vector (on an FT system).
Test vectors should be selected for larger circuit blocks to provide good logic coverage. For example, an $n$-qubit adder often is designed to execute $\vert a\rangle\vert b\rangle\vert0\rangle \rightarrow \vert a\rangle\vert a+b\bmod 2^n\rangle\vert0\rangle$ (where the last register is ancillae that must be cleaned)~\cite{draper2000addition}. We might choose to confirm that adding zero, adding one, carrying within a register, and carry out of the register all work properly. Thus, a reasonable set of test vectors for $n=4$ might be $\vert0000\rangle\vert1111\rangle\vert0\rangle$, $\vert0001\rangle\vert0011\rangle\vert0\rangle$, and $\vert0001\rangle\vert1111\rangle\vert0\rangle$.

\subsection{Testing and debugging Phase Modulation (PM) blocks}

Testing and debugging AP blocks were relatively straightforward; however, moving on to the PM and AR blocks gets more challenging. This subsection will consider a strategy for debugging a PM block. The most straightforward PM block would be a circuit containing any phase gate, such as a T, Z, CZ, or S gate.

The challenge in creating test vectors for PM blocks is choosing the correct tests to detect subtle phase shifts. These shifts, often representing the core computational output of the circuit, require high precision in both the generation and measurement phases. The approach typically involves initializing the circuit in a superposition state where the effects of phase modulation can be maximally observed. Cirquo's \texttt{fQuantTester(circuit, test\_vectors)} allows users to use their test vectors as state vectors to perform unit testing on PM and AR blocks.

Though the test in Figure~\ref{simtst} can tell us whether an error in the circuit exists, it can not provide further information about that error. However, we can use other approaches to get more information about the possible error.

The strategy we propose here is to use the swap test~\cite{foulds2021controlled}. 
The swap test is a fundamental quantum computing procedure used to determine the similarity between two quantum states. It is beneficial for measuring the inner product of two states, which can then be utilized to calculate their fidelity or similarity.

Consider two quantum states \( \vert\psi\rangle \) and \( \vert\phi\rangle \) that we want to compare. The swap test involves an ancillary qubit (the control qubit) initialized in the state \(\vert0\rangle\) and the two states \( \vert\psi\rangle \) and \( \vert\phi\rangle \). The process of the swap test involves the following steps:

\begin{enumerate}
    \item Apply a Hadamard gate to the control qubit, putting it into the superposition \( \frac{1}{\sqrt{2}}(\vert0\rangle + \vert1\rangle)\).
    \item Perform a CSWAP gate using the control qubit. The CSWAP gate swaps \(\vert\psi\rangle\) and \(\vert\phi\rangle\) only if the control qubit is in the state \(\vert1\rangle\).
    \item Apply another Hadamard gate to the control qubit.
    \item Measure the control qubit. The probability of measuring \( \vert0\rangle\) is given by \(P(0) = \frac{1}{2} + \frac{1}{2}\vert\langle\psi\vert\phi\rangle\vert^2\).
\end{enumerate}

The outcome of the measurement gives us information about the similarity of the two states. If the states are identical, the probability of measuring \( \vert0\rangle \) in the control qubit will be 1. If they are orthogonal, the probability will be \(\frac{1}{2}\).
 
We can further analyze the swap test results for more information about the phase difference. To do that, consider \(\vert\psi\rangle = \vert0\rangle + e^{i\theta_1}\vert1\rangle\) and \(\vert\phi\rangle = \vert0\rangle + e^{i\theta_2}\vert1\rangle\). We can then recalculate the probability of measuring $1$ or $0$ as a function of \(\theta_1\) and \(\theta_2\) as follows:

The inner product of these two states is
\begin{equation}
\langle\psi\vert\phi\rangle = \frac{1}{2}\left(1 + e^{i(\theta_2 - \theta_1)}\right).
\end{equation}

We need to find the magnitude squared of this inner product. Since \( e^{i(\theta_2 - \theta_1)} \) can be expressed as \( \cos(\theta_2 - \theta_1) + i\sin(\theta_2 - \theta_1) \),  and \(\Delta\theta = \theta_2 - \theta_1\) we get
\begin{equation}
\left\vert \frac{1}{2}\left(1 + e^{i(\Delta\theta)}\right) \right\vert^2 = \frac{1}{4}\left( \left[1 + \cos(\Delta\theta)\right]^2 + \left[\sin(\Delta\theta)\right]^2 \right).
\end{equation}

The probability \( P(0) \) is given by
\begin{equation}
P(0) = \frac{1}{2} + \frac{1}{2}\left\vert\langle\psi\vert\phi\rangle\right\vert^2.
\end{equation}
Substituting the magnitude squared into this, we get

\begin{align}
  P(0) &= \frac{1}{2} + \frac{1}{8}\left( \left[1 + \cos(\Delta\theta)\right]^2 + \left[\sin(\Delta\theta)\right]^2 \right)\\
 &= \frac{3}{4} + \frac{1}{4} \cos (\Delta\theta).
\end{align}

Since \( P(1) = 1 - P(0) \), we substitute our expression for \( P(0) \) and simplify
\begin{equation}
P(1) = \frac{1}{4} - \frac{1}{4}\cos(\Delta\theta).
\end{equation}

After implementing the swap test in Qiskit (Listing~\ref{swapppp}), we calculate the squared inner product using the number of shots using the equation \(s = 1 - \frac{2}{N} B\). $s$ is the squared inner product, $N$ is the total number of shots, and $B$ is the number of times $1$ was measured.

\begin{lstlisting}[language=Python, caption=An Implementation of a simple swap test using Qiskit., label=swapppp]
q = QuantumRegister(3, 'q')
c = ClassicalRegister(1, 'c')
circuit = QuantumCircuit(q, c)

circuit.h(q[0]) 
circuit.cswap(q[0], q[1], q[2])  
circuit.h(q[0])

circuit.measure(q[0], c[0]) 

simulator = Aer.get_backend('qasm_simulator')
N = 8192
job = execute(circuit, simulator, shots=nShots)
counts = job.result().get_counts()

if '1' in counts:
    B = counts['1']
else:
    B = 0
s = 1 - (2 / N) * B
\end{lstlisting}

Since \( P(1) = 1 - P(0) \) and \( \frac{B}{N} = P(1) \), we have:
\begin{equation}
\frac{B}{N} = \frac{3}{4} + \frac{1}{4} \cos (\Delta\theta).
\end{equation}

\begin{figure}
\centering
\includegraphics[width=\columnwidth]{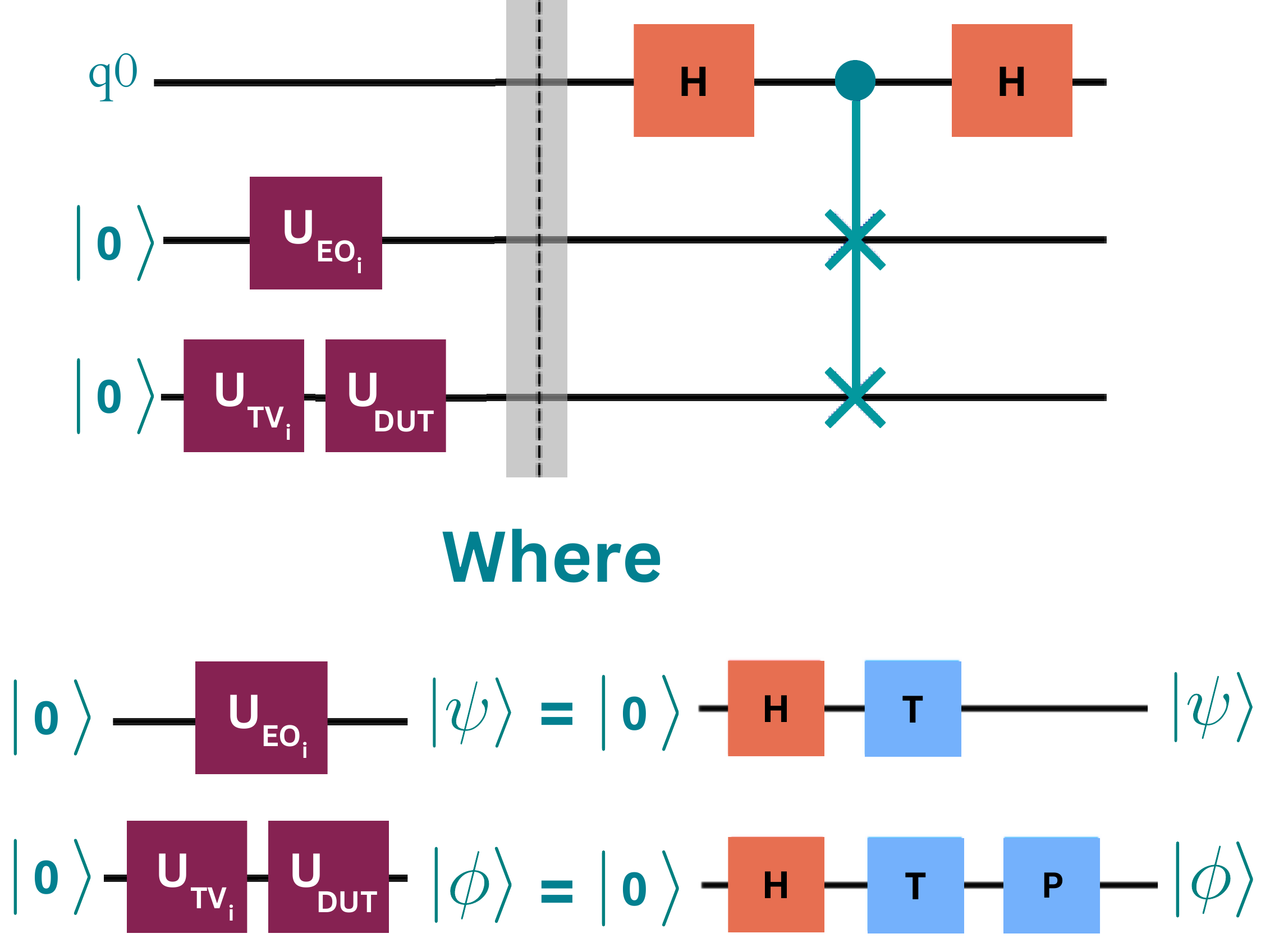}
\caption{Using the swap test to detect the phase difference in a circuit containing a phase error $p$. $U_{DUT}$ is the Device Under Test (the full circuit/slice). $U_{TVi}$ is the circuit corresponding to test vector $i$. $U_{EOi}$ is the expected behaviour for $TV_{i}$.}
\label{swaperr}
\end{figure}

To see this better, assume we have the circuit shown in Figure~\ref{swaperr}. 

In testing the circuit, the programmer will first use the technique in Figure~\ref{simtst}. Assuming that test returns something other than $\vert0\rangle$, meaning $U_{DUT}U_{TVi}\ne U_{EOi}^\dagger$, the next step is to attempt to isolate the bug.  It is useful to find the angle of the extra (or missing) angle, as well as to find the qubit(s) with the extra gate.

We can denote that phase error as a gate P applied to $\vert\phi\rangle$ then use the swap test to determine that phase.
To do that we use Cirquo's \texttt{applySwapTest(circuit, k, reg1\_index, reg2\_index, nshots)}, with \textit{nshots = 8192}. In this example, let us say that we found $s = 0.76$. The function will also return the value of $\Delta\theta$ calculated according to \(\vert\langle\psi\vert\phi\rangle\vert^2 = \frac{1}{2} + \frac{1}{2}\cos(\Delta\theta)\).
\begin{align}
0.76 &= \frac{1}{2} + \frac{1}{2} \cos (\Delta\theta)\\
\frac{1}{2} &= \cos (\Delta\theta).    
\end{align}

In our example, we get \(\Delta\theta = \frac{\pi}{3}\). From that we can infer the presence of an extraneous $Z(\frac{\pi}{3})$ gate. \\

\begin{figure}
\centering
\includegraphics[scale=0.25]{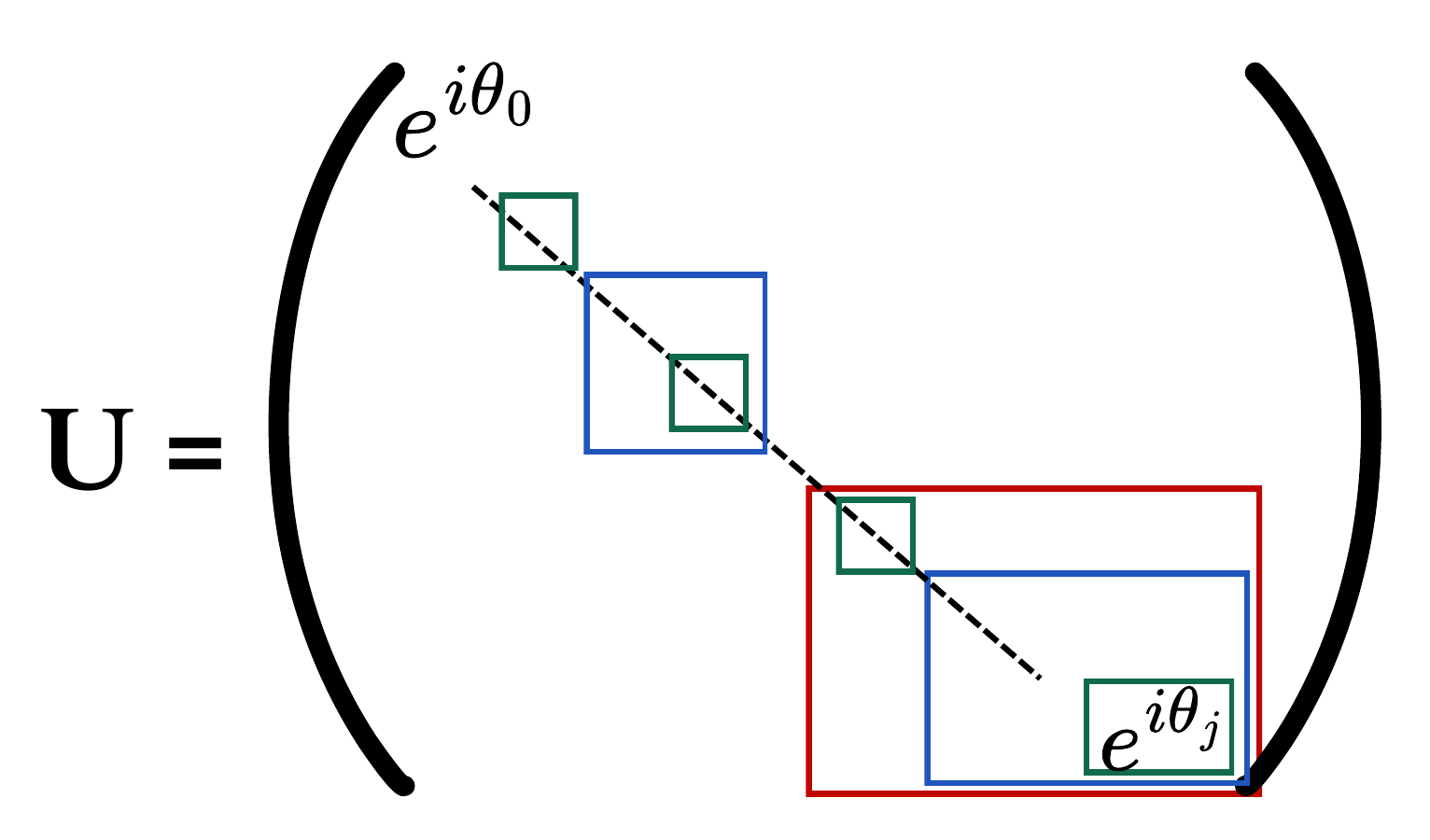}
\caption{The unitary for a three-qubit phase slice. The colored boxes delineate terms affected by errors in the high-order (red), middle (blue), and low-order (green) qubits.}
\label{ppp}
\end{figure}

Two common errors will be (a) applying the correct phase shift to the wrong qubit, and (b) applying the wrong phase shift to the right qubit (including omitting a needed gate). (For simplicity, we omit controlled phase gates here.) Test vectors need to be prepared to help us distinguish these two cases. For example, the unitary for a three-qubit phase slice is shown in Figure~\ref{ppp}.  

A phase gate on the high-order qubit will be reflected in the values in the red box, the middle qubit in the blue boxes, and the low-order qubit in the green boxes. Testing the input $\vert000\rangle + \vert111\rangle$ will tell us if the sum of the phase shifts on all three qubits is correct. If it is not correct, enough shots of the test can tell us the incorrect angle but do not tell us where to look. For independent, single-qubit gates, testing the cases $\vert000\rangle+\vert001\rangle$, $\vert000\rangle+\vert010\rangle$, and $\vert000\rangle+\vert100\rangle$ can isolate the qubit with the error. If the slice contains controlled phase gates, we may need a larger set of test vectors, e.g., $\vert110\rangle+\vert111\rangle$. With careful test vector design, we can test a small number of cases, hopefully, $O(n)$ or $\Tilde{O}(n)$ to develop high confidence in the behavior of the entire slice, which has $2^n$ angles $\theta_{i}$, each potentially different.\\


\subsection{Testing and debugging Amplitude Redistribution (AR) blocks}

Testing and debugging AR blocks can be quite challenging. However, once we differentiate between PM and AR blocks, we can use different approaches to test and debug them. We just discussed how the swap test could be utilized to debug PM blocks. Though we can still use the swap test or the approach in Figure~\ref{simtst} to validate the results from AR blocks, they will not provide much information that we can use except for the degree of difference between the expected and resultant states.

In most cases, AR blocks will be functions offered by the package the user decides to use. Hence, they often will not need to build it from scratch. However, it is essential to address how developers can create helpful test vectors for AR blocks.

The primary challenge here is the exponential increase in complexity with the number of qubits, which makes exhaustive testing impractical for larger circuits. AP and PM circuits allowed us to work with single amplitudes for each test, but AR circuits depend on interference, and can change the number of non-zero amplitudes, substantially complicating tests, especially in terms of interpreting test outputs. Additionally, since amplitude redistribution can result in highly entangled states, ensuring that the generated test vectors adequately reflect the potential entanglement patterns is crucial for thorough testing. 

Simulation, induction or inference from smaller cases, and sampling techniques to assess distributions become essential, with the aim of capturing the most significant aspects of amplitude redistribution with a manageable number of test vectors. 

\begin{figure}
\centering
\includegraphics[width=\columnwidth]{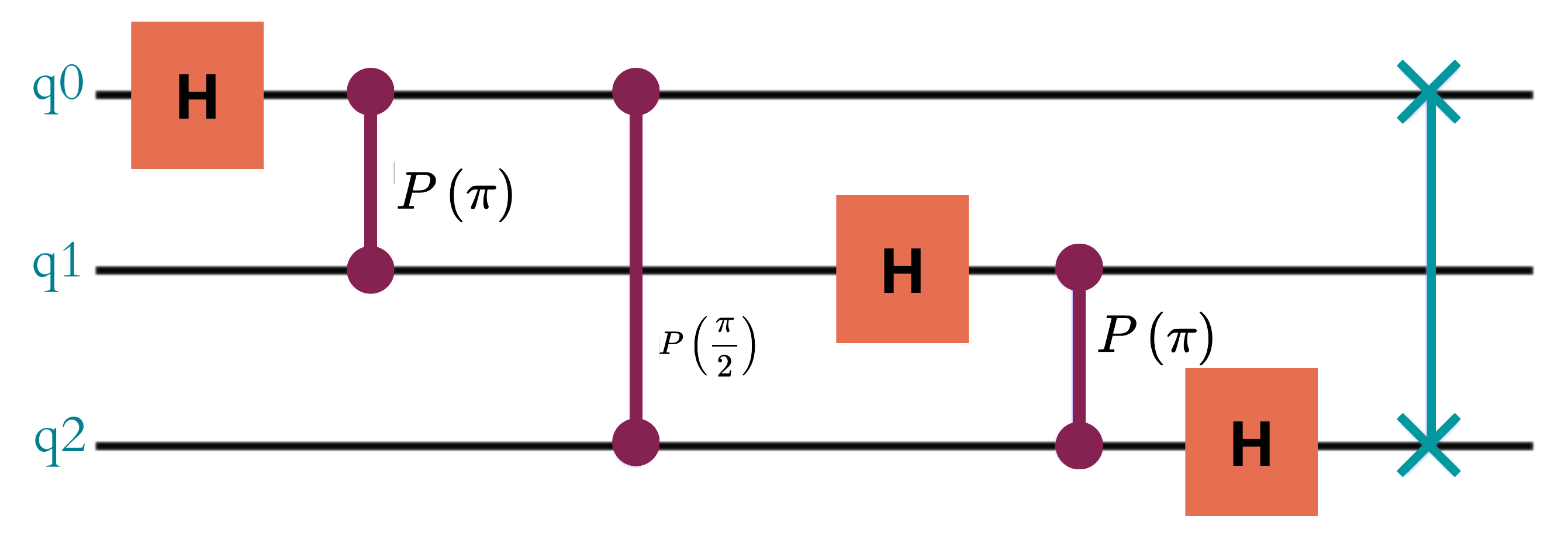}
\caption{A 3-qubit QFT circuit.}
\label{qftc}
\end{figure}

Consider the Quantum Fourier Transform (QFT), a fundamental operation in quantum computing, pivotal for algorithms like Shor's algorithm~\cite{shor1999polynomial} for factoring and quantum phase estimation. Ensuring its correct implementation is vital for the success of these quantum algorithms. The QFT on an \(n\)-qubit state is defined as
\begin{equation}
    QFT\vert j\rangle = \frac{1}{\sqrt{2^n}} \sum_{k=0}^{2^n - 1} e^{2\pi ijk / 2^n} \vert k\rangle.
\end{equation}

A general implementation using Qiskit can be seen in Listing~\ref{qftcode}.  The code here offers an important clue to debugging AR circuits: the code is parameterized in $n$, the number of qubits in the QFT. Thus, the programmer can test and debug their code using small values of $n$ and develop confidence in the behavior of larger instances. We can use simulators as well as Cirquo's functions for examining the code to test propositions about the circuit.

\begin{lstlisting}[language=Python, caption=An Implementation of the QFT using Qiskit., label=qftcode]
def qft(circuit, n):
    """Applies QFT on the first n qubits in circuit"""
    for j in range(n):
        circuit.h(j)
        for k in range(j+1, n):
            circuit.cp(2 * pi / (2**(k - j)), k, j)
    for j in range(n//2):
        circuit.swap(j, n-j-1)
\end{lstlisting}

One approach to creating efficient test vectors is to focus on the properties of the QFT, such as linear shift-invariance and parallelism.

In this example, we will focus on the linear shift-invariant aspect of the QFT. This result of applying the QFT on a periodic state shows the periodicity and phase relations encoded by the original state transformed into a new superposition reflecting these frequency domain properties. A periodic $n$-qubit state ($\vert\psi(n, r,l)\rangle$) is a state that shows a periodic behavior and can be defined by its period $r$ and its shift $l$. We can express that as
\begin{equation}
    \vert\psi(n, r, l)\rangle = \sum_{k=0}^{2^n - 1} c_k \vert k\rangle
\end{equation}
where $c_k$ are the coefficients of the computational basis states $\vert k\rangle$, and they reflect the periodicity and shift of the state. Specifically, $c_k$ is non-zero for states $\vert k\rangle$ that satisfy \((k - l) \mod r = 0\).
For example, consider $\vert\psi(3, 2,1)\rangle$
\begin{equation}
 \vert\psi(3, 2,1)\rangle = \frac{1}{\sqrt{N}} \left( \vert001\rangle + \vert011\rangle + \vert101\rangle + \vert111\rangle \right).   
\end{equation}

We can now apply the QFT to state $\vert\psi(3, 2,1)\rangle$ and observe how it preserves the period while transforming it to the frequency domain.
For a 3-qubit system (\(n=3\)), the QFT equation simplifies to
\begin{equation}
    QFT\vert j\rangle = \frac{1}{\sqrt{8}} \sum_{k=0}^{7} e^{2\pi ijk / 8} \vert k\rangle.
\end{equation}

We can see the circuit for a 3-qubit QFT in Figure~\ref{qftc}. Applying the QFT to \(\vert\psi(2,1)\rangle\) gives
\begin{align}
    \vert\psi_{QFT}\rangle = \frac{1}{2\sqrt{8}} \Bigg( &\sum_{k=0}^{7} e^{2\pi i \cdot 1 \cdot k / 8} \vert k\rangle + \sum_{k=0}^{7} e^{2\pi i \cdot 3 \cdot k / 8} \vert k\rangle \nonumber \\
    + &\sum_{k=0}^{7} e^{2\pi i \cdot 5 \cdot k / 8} \vert k\rangle + \sum_{k=0}^{7} e^{2\pi i \cdot 7 \cdot k / 8} \vert k\rangle \Bigg).
\end{align}

After simplifying this equation, we get
\begin{equation}
    \vert\psi_{QFT}\rangle = \frac{1}{\sqrt{2}} (\vert0\rangle - \vert2\rangle + \vert4\rangle - \vert6\rangle).
\end{equation}

The calculation demonstrates how the QFT reveals the periodicity and phase relationships encoded in the original state \(\vert\psi(n, r, l)\rangle\), transforming it into a new superposition that reflects these properties in the frequency domain. That effect can be seen in Figure~\ref{qftc}-A.

\begin{figure}
\centering
\includegraphics[scale=0.4]{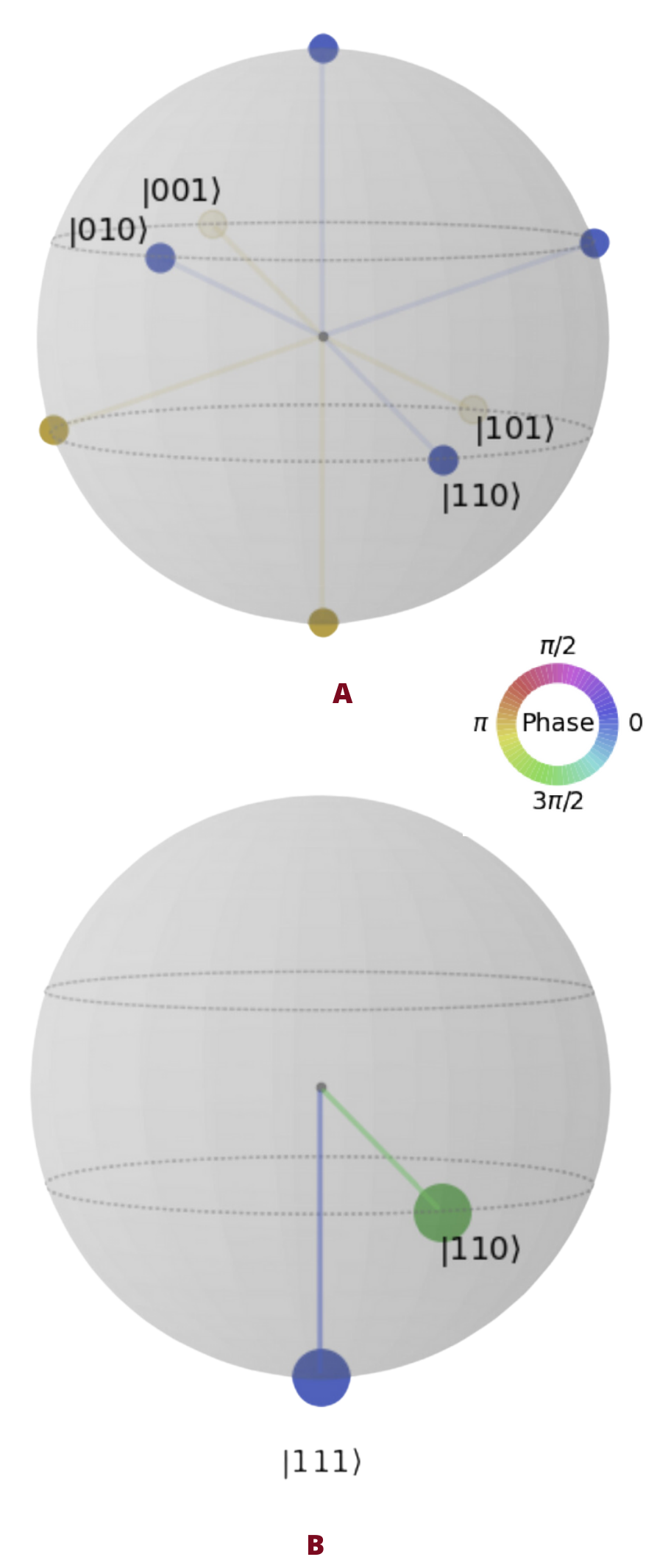}
\caption{(A) The Q-sphere results for the periodic state \(\vert\psi(3, 2, 1)\rangle\). (B) The Q-sphere results of the 3-qubit QFT where a Hadamard gate on qubit $0$ is missing. }
\label{qfte}
\end{figure}

Consider the QFT for $n=1024$; full simulation of this circuit is well beyond classical capabilities. For a fault-tolerant machine, we expect to use it frequently. In Shor's algorithm, the QFT is far less expensive than the modular exponentiation portion of the circuit, and so the execution cost will be reasonable.  As a first test, the programmer can run the full QFT circuit on a chosen test vector or set of test vectors using the approach of Figure~\ref{simtst}. If this test returns an error, the next step is to reduce the scale of the tests to something that can be simulated.

Assume that the user forgot to apply a Hadamard gate to qubit $0$ at the beginning of the algorithm's implementation. (This kind of error can be expected to be common, as a result of off-by-one programming mistakes in loops, as in Listing~\ref{qftcode}.) Simulating the circuit for \(\vert\psi(3, 2, 1)\rangle\) will lead to the results shown in Figure~\ref{qfte}-B. We can see that the output does not match our expected results but we do not know what might be causing the error. 

Because the most common error when implementing quantum algorithms is often a missing/extra gate, we can get the gates count using Qiskit's \texttt{count\_ops()}. However, this will only return the total count, including the measurement, which can be helpful to use with Cirquo's \texttt{gateLoc()} to get the specific qubit and line of code where a gate was added.

We know that an $n$-qubit QFT includes $n$ Hadamard gates, $\frac{n(n-1)}{2}$ controlled phase rotations are required, and $\left\lfloor\frac{n}{2}\right\rfloor$ SWAP gates. 

These two functions allow us to discover that qubit $0$ has no Hadamard gates applied; adding one (or correcting the loop conditions) and executing the circuit again fixes the error. Repeating the smaller tests up to the limit of simulation, followed by re-testing the full circuit on the FT computer, increases our confidence in the overall circuit.

In the prior subsections, we worked with single amplitudes in test vectors. Because AR blocks modify the set of amplitudes, the output of a single-amplitude input will be a multiple-amplitude output. With many amplitudes, testing results will be stochastic, and reconstructing the amplitude distribution will be difficult.

An alternative would be to engineer an input test vector to generate a single amplitude on output.  We could do this, for example, by taking a desired output, directly calculating the inverse of the AR block, and using the output of that as the initial input test vector.  Unfortunately, in many cases, the direct creation of that input vector will be as complicated as the full AR block.  However, the approaches shown in Figures~\ref{simtst}  and \ref{swaperr} can be used, for example, to confirm that a new implementation of the QFT produces the same output as a standard library implementation. Working entirely from scratch will be difficult, but once bootstrapped comparing versions with new optimizations or features can be more straightforward using these techniques.



\section{Discussion}
\label{sec:dis}

One significant challenge in quantum program testing is the complexity of quantum programs. Unlike classical programs, quantum programs often involve entangled states and superpositions, making isolating and identifying bugs difficult. Techniques from classical debugging, such as binary search debugging, can be extended and adapted for quantum computing. This involves dividing the quantum program into smaller sections, testing them individually, and eliminating those without bugs~\cite{srivastva2017debugging, arulraj13apollo, di2021s}. Though that can be beneficial in some cases, quantum-specific challenges, such as cross-register entanglement, require novel solutions.

\subsection{Related Work}
Some classical approaches can be adapted on the quantum side, such as "slicing." Quantum circuits can be sliced horizontally and vertically into smaller sections that are easier to test and debug. This, however, is not as straightforward as the classical slicing, as we need to address the entanglement within the circuit. This can be done using the Kronecker product as discussed by CutQC~\cite{tang2021cutqc}.
Other approaches include hypothesis testing, a statistical method adapted to observe and analyze the behavior of quantum circuits under different inputs~\cite{liu2006statistical, mccauley2008debugging}. In addition to runtime testing and debugging of quantum circuits using statistical methods~\cite{huang2019statistical, weder2020quantum, miranskyy2021testing, rand2018qwire}.

The most complete quantum development tool is Microsoft's Q\#, which offers unit testing functionality for quantum circuits for NISQ-era quantum program~\cite{Li}. However, writing efficient test cases for quantum circuits can be challenging, especially if the type of circuit is unknown. Q\# recently added a visual quantum resource estimation so users can track the execution of their programs on the hardware. 

IBM has made a similar effort with Qiskit's Trebugger~\cite{qiskit_trebugger_2022}. The Trebugger is a debugger for the transpiler offered by Qiskit. Though both the Q\# resource estimation and Qiskit's Trebugger can be used to find errors in the code, both aim to make the circuit optimization process more efficient for specific hardware and not to debug the software.

\subsection{Challenges and Limitations}
At a high level, debugging quantum circuits is essentially performing process tomography. We have an expectation of how the state of the system evolves over time, and the goal of our testing is to determine if the system is following these expectations. 
In Section~\ref{sec:test}, we proposed different strategies to use when testing and debugging the various types of blocks, such as using the swap test for PM blocks or targeting the algorithm's properties to create test vectors for AR circuits. Though these approaches are theoretically valid to test and debug these circuits, they still have limitations. 

\subsubsection{Creating useful test vectors}
Testing AR and PM blocks such as the Quantum Fourier Transform (QFT) poses unique challenges. The probabilistic outcomes of quantum states require statistical methods for validation, diverging from the deterministic testing used in classical computing. Entanglement further complicates testing, as the state of one qubit is interdependent with others, obscuring individual analysis. 
One way to approach this is by creating test vectors that prove the properties of the block under test, similar to what we did in Section~\ref{sec:test}. Using basic test vectors (applying the block for different bases) can provide the user with information to assist them in the debugging process.

\subsubsection{The increase in needed resources}
The swap test circuit looks fairly simple, but its complexity is directly tied to the size of the states being compared. For two quantum states, each consisting of \(n\) qubits, the circuit complexity \(C_{\text{swap test}} = n\), where \(C_{\text{swap test}}\) denotes the number of controlled-SWAP gates needed. 

Each CSWAP gate has a significant cost in terms of quantum resources. Implementing a CSWAP gate typically involves multiple elementary quantum gates, such as CNOT and single-qubit gates, depending on the architecture of the quantum processor. 

Therefore, the resource cost and potential error rates increase with the number of CSWAP gates used. As \(n\) grows, the swap test circuit becomes more complex and more challenging to execute accurately due to increased gate operations and the associated error rates. 

A similar argument can be made for implementing and executing process tomography (QPT)~\cite{beck2012process, mohseni2008quantum}.

QPT aims to reconstruct the complete quantum process, represented by a superoperator acting on a density matrix. The complexity of QPT scales exponentially with the number of qubits \(n\) in the system~\cite{zambrano2020estimation}. The number of experimental configurations required for full reconstruction grows~\cite{torlai2023quantum} is

\begin{equation}
    C_{\text{QPT}} = 4^n
\end{equation}

In contrast, Selective Process Tomography (SQPT)~\cite{bendersky2009selective, bendersky2008selective} focuses on reconstructing only specific elements of the process matrix. The number of experimental configurations in SQPT depends on the number of elements being selectively measured. This selective approach reduces the scaling, often to a polynomial relation with the number of qubits, depending on the targeted elements. Thus, for a subset of elements \(k\)~\cite{gaikwad2022implementing}, the scaling is

\begin{equation}
    C_{\text{SQPT}} \approx \text{poly}(n, k)
\end{equation}

The difference in scaling between QPT and SQPT has significant implications for their applicability in large quantum systems.

\subsubsection{Estimating the optimal number of shots needed}
Estimating the required number of shots in quantum circuit simulations depends on the type of circuit we are testing/ executing. For the AP blocks, we only need one shot to get the results, however, that is not the case when we examine PM and AR blocks. These blocks need more shots to have a higher accuracy. The swap test, for example, involves considerations about statistical accuracy, computational resources, and the specifics of the quantum states being compared. 
The relation between the number of shots and the standard deviation ($\sigma$) of the measurement outcome can be described by
\begin{equation}
    \sigma = \sqrt{\frac{p(1-p)}{N}}
\end{equation}
where $p$ is the probability of measuring a particular outcome, and $N$ is the number of shots.
While there is no simple formula to calculate the optimal number directly, we can determine a suitable number of shots based on the requirements for statistical accuracy. We can use the $\sigma$ with a specific accuracy value $Z$ to calculate the confidence interval (CI) of the number of shots needed using the formula
\begin{equation}
CI = p \pm Z \times \sigma
\end{equation}

\subsubsection{Distinguishing between hardware and software errors}
A critical aspect of quantum program testing is acknowledging the differences in debugging on simulators, NISQ machines, and fault-tolerant quantum computers. Limitations in program size and hardware-related errors in NISQ machines pose significant challenges. While these issues might be resolved with fault-tolerant quantum computers, classical computers will still play a vital role in debugging and testing quantum programs.

Developing a deep understanding of the behavior of the circuits, identifying sources of errors, and creating effective test cases are essential for building the quantum intuition necessary to advance the field toward the fault-tolerant era.

\section{Conclusion}
\label{sec:conc}

Testing and debugging quantum programs represent both an opportunity and an obstacle in quantum computing, filled with unique challenges and limitations. While significant progress has been made in developing methodologies like Quantum Process Tomography and Selective Process Tomography, the inherent complexities of quantum mechanics continue to pose substantial obstacles.

One of the primary challenges is the exponential scaling of system complexity with the number of qubits. This not only makes the execution of quantum programs more resource-intensive but also exponentially increases the difficulty of debugging and error correction. Furthermore, the no-cloning theorem and the probabilistic nature of quantum computing add complexity to debugging, as they prevent the straightforward replication and observation of quantum states.

Additionally, the current quantum hardware stage, often called NISQ devices, introduces practical limitations. These devices are prone to errors and have limited coherence times, which restrict the reliability and scalability of quantum programs. Implementing effective error correction and noise mitigation strategies thus remains a crucial area of ongoing research.


Despite these challenges, the field of quantum computing holds immense potential. The development of more sophisticated testing and debugging tools and advancements in quantum hardware will pave the way for more robust and reliable quantum programs. As the field continues to mature, it is expected that many of the current limitations will be overcome, unlocking the full potential of quantum computing in solving complex problems beyond the reach of classical computers.

In summary, the journey towards fully harnessing the power of quantum computing is just beginning. The challenges and limitations in testing and debugging quantum programs are significant, yet they provide a fertile ground for innovation and discovery in the field.


\section*{Acknowledgements}
We would like to thank Michal Hajdušek and Naphan Benchasattabuse for their invaluable input to this work.

\bibliographystyle{IEEEtran}
\bibliography{bib}

\appendices
\section{Cirquo's Functions}
\label{app1}

Cirquo is built using Python on top of the Qiskit module. In Qiskit, any quantum circuit is built using the object class QuantumCircuit. Any QuantumCircuit object can contain QuantumRegisters, ClassicalRegisters, different quantum gates, and measurement operations. The Qiskit QuantumCircuit object contains many properties and characteristics. In order to build our tool, we extended this class to include a few new commands to include breakbarriers (the quantum equivalent of breakpoints) to cut the circuit and gate-tracking options. In addition, we added new functionalities to perform horizontal and vertical slicing.~\cite{padmanabhan1986uncertainty}. 

The methods added to the QuantumCircuit class and the new functions offered by Cirquo are:
\begin{enumerate}
  \item {\tt breakbarrier()}: a new object type based on Qiskit's barrier class that is used to pinpoint where the tool is going to cut the circuit when using the Vertical tool function (\textbf{VSlicer}).
  \item {\tt gateInfo()}: a method that is used to store information about all gates added to the circuit when the debugging mode is enabled. The information is the gate type, the number of occurrences, and where this gate was added to the circuit in the code.
  \item {\tt startDebug()}: This function enables the debugging mode by extending the QuantumCircuit Class to include both \emph{breakbarrier} and \emph{gateInfo} methods.
  \item {\tt endDebug()}: This function disables the debugging mode.
  \item {\tt VSlicer()}: This function takes a QuantumCircuit object that contains breakbarriers and then divides the circuit based on the location of those \emph{breakbarrier} and returns the original circuit as well as a list of subcircuits corresponding to the circuit dividing based on the \emph{breakbarrier} locations.
  \item {\tt HSlicer()}: This function removes unused qubits or QuantumRegisters from a subcircuit after using the vertical slicer.
  \item {\tt gateLoc()}:  This function takes a circuit or a subcircuit, a gate, and a qubit or a list of qubits (optional), and displays how many times and where in the code this gate was added to the circuit.
  \item {\tt catCircuit()}:  This function takes a circuit and categorizes it into either AP, PM, or AR based on its size and the containing gates.
  \item {\tt pClassAnalyzer()}:  This function is used to run tests on AP circuits to observe the circuit's behavior.
  \item {\tt fQuantAnalyzer()}:  This function is used to run tests on PM and AR circuits to observe the circuit behavior.
  \item {\tt pClassTester()}:  This function is used to test AP blocks.
  \item {\tt fQuantTester()}:  A function to run tests on PM and AR blocks.
  \item {\tt applySwapTest()}: A function to apply the swap test and calculates $\delta\theta$ for PM blocks.
\end{enumerate}

\section{Reading a Q-sphere}
\label{app2}

The Q-sphere~\cite{qsphere} is an approach introduced by IBM to represent the state of a system of one or more qubits by associating each computational basis state with a node on the surface of a sphere. Each node's radius is proportional to the probability of its basis state. The node color indicates its phase according to the phase color circle in the bottom right corner of the Q-sphere figure.

\begin{figure}
\centering
\includegraphics[scale=0.2]{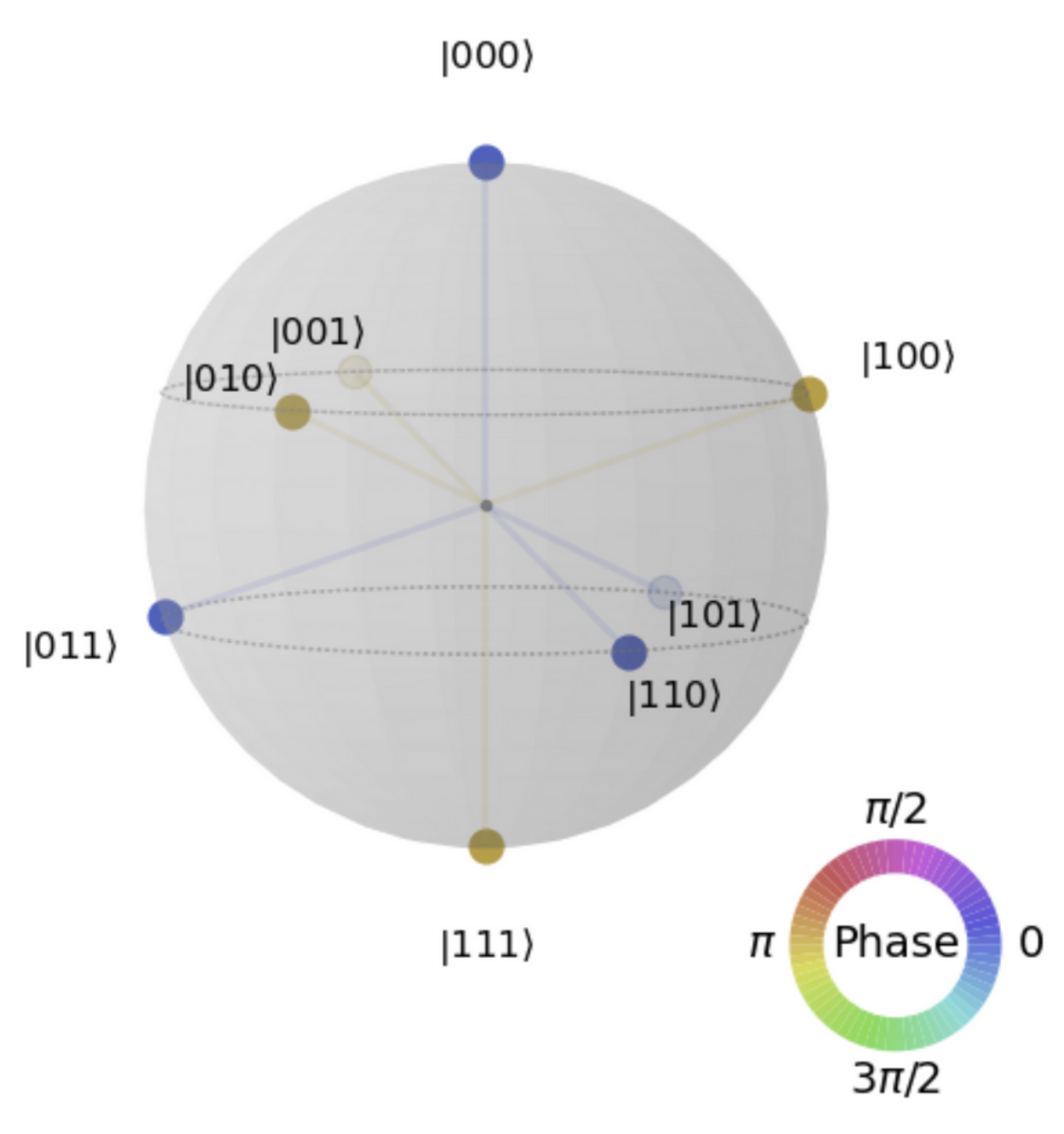}
\caption{The Q-sphere representation of $\vert000\rangle-\vert001\rangle-\vert010\rangle+\vert011\rangle-\vert100\rangle+\vert101\rangle+\vert110\rangle-\vert111\rangle$.}
\label{qexp}
\end{figure}

In a Q-sphere, we place the state where all qubits are 0 at the sphere's top (north pole) and where all qubits are $1$ at the bottom (south pole). Other states are arranged in between, forming circles of latitude based on their Hamming distance from the $\vert0\rangle$ state. For example, if we have a circuit constructing a system of 3 qubits in superposition, except that states $\vert001\rangle, \vert010\rangle, \vert100\rangle$, and $\vert111\rangle$ have phase equal to $\pi$. The Q-sphere visualization of that system can be seen in Figure~\ref{qexp}.

\EOD

\end{document}